\documentclass[prl,aps,floatfix, superscriptaddress]{revtex4}
\usepackage{amsmath,bm,graphicx}
\usepackage{amssymb}
\usepackage{amsfonts}
\usepackage{color}

\newcommand{\tb}{\textcolor{black}}

\begin{document}
\title{Rectification and non-Gaussian diffusion in heterogeneous media}

\author{P. Malgaretti}
\email[Corresponding Author : ]{malgaretti@is.mpg.de }
\affiliation{Max-Planck-Institut f\"{u}r Intelligente Systeme, Heisenbergstr. 3, D-70569
Stuttgart, Germany}
\affiliation{IV. Institut f\"ur Theoretische Physik, Universit\"{a}t Stuttgart,
Pfaffenwaldring 57, D-70569 Stuttgart, Germany}
\affiliation{Departament de Fisica de la Mat\`eria Condensada, Facultat de Fisica,  Universitat de Barcelona, carre Mart\'i i Franques 1, Barcelona 08028, Spain}
\author{I. Pagonabarraga}
\affiliation{Departament de Fisica de la Mat\`eria Condensada, Facultat de Fisica,  Universitat de Barcelona, carre Mart\'i i Franques 1, Barcelona 08028, Spain}
\author{J. M. Rubi}
\affiliation{Departament de Fisica de la Mat\`eria Condensada, Facultat de Fisica,  Universitat de Barcelona, carre Mart\'i i Franques 1, Barcelona 08028, Spain}



\begin{abstract}

We show that when Brownian motion takes place in a  heterogeneous medium, the presence of local forces and transport coefficients leads to deviations from a Gaussian probability distribution that make that the ratio between forward and backward probabilities depends on the nature of the host medium, on local forces and also on time. We have applied our results to two situations: diffusion in a disordered medium and diffusion in a confined system. 
\tb{For such scenarios we have shown that our theoretical predictions are in very good agreement with numerical results. Moreover we have shown that the deviations from the Gaussian solution lead to the onset of rectification.} Our predictions could be used to detect the presence of local forces and to characterize the intrinsic short-scale properties of the host medium, a problem of current interest in the study of micro and nano-systems.
\end{abstract}

\maketitle

\section{Introduction}

The symmetry of the probability distribution of a system in equilibrium, expressed through the detailed balance condition  breaks down when a driving force is applied~\cite{Hanggi_RMP}. The ratio of probabilities between forward and backward  particle displacements  is in this case independent of time, equal to a Boltzmann factor. For a Brownian particle under a constant conservative force, $f_0$, such as a gravitational~\cite{Astumian_Am_Phys}, optical~\cite{Astumian_JCP} or entropic~\cite{david-miguel} force,  the   ratio is given by:
\begin{equation}
 \frac{p(\Delta x,t|x_0,t_0)}{p(-\Delta x,t|x_0,t_0)}=e^{\beta f_0 \Delta x}
\label{cnst-force}
\end{equation}
where $p(\Delta x,t|x_0,t_0)$ is the probability of measuring a particle displacement of magnitude $\Delta x$ at time $t$, given the initial condition $p(x_0,t_0)$ and $\beta^{-1}=k_B T$ with $k_B$ the Boltzmann constant. Eq.(\ref{cnst-force}) has been obtained using different theoretical frameworks~\cite{Hanggi_RMP,Ciliberto} and for different observables such as entropy production rate~\cite{Gallavotti} or  mechanical work~\cite{Crooks}.

The peculiar form of the ratio between probabilities  given by Eq.(\ref{cnst-force}) is a consequence of the Gaussian nature of the probability distribution function (pdf)~\cite{Astumian_JCP,david-miguel}, solution of the corresponding Smoluchowski equation, and of the potential nature of the force~\cite{Vainstein}. For a $1D$ dynamics, as is the case of Eq.(\ref{cnst-force}), forces are always potential ensuring a Gaussian probability distribution function. For a $3D$ dynamics, as in the case of Brownian motion in a shear flow, the ratio between probabilities depends on time due to the fact that the shear flow is not potential~\cite{Vainstein}.
In a variety of  situations, such as for particles diffusing in porous media or displacing through ion channels or membrane pores, the assumption of a constant force and/or transport coefficients is not justified. For these local transport scenarios Eq.(\ref{cnst-force}) cannot be applied.

In this work, we will characterize the dynamics of Brownian particles when they displace in an heterogeneous environment in which transport coefficients and forces may depend on position. The local  nature of these quantities leads to a non-Gaussian pdf for particle displacement due to a coupling between particle advection and diffusion. This behavior has not been reported in previously studied systems, where particles diffuse in a homogeneous medium and are subjected to uniform forces~\cite{Astumian_Am_Phys,Astumian_JCP,david-miguel}~\footnote{A case of nonuniform forces where an analytical solution of the associated Smoluchowski equation exists is the Ornstein-Uhlenbeck process. However, in this case the probability distribution converges rapidly to a Gaussian due to the confining nature of the potential~\cite{Risken}.}.
Hence, for local transport, the time evolution of the particle pdf cannot be regarded as a diffusion process with respect to  a moving mean value, rather convection and diffusion affect each other non-trivially leading to the appearance of new dynamical regimes. \tb{In particular our results highlight the presence of a rectification regime in which particle transport benefits from the heterogeneity of the medium}.

The article is organized as follows. In Section 2, we derive the equivalent of Eq.~(\ref{cnst-force})  for local transport, in which the diffusion coefficient and the driving forces may depend on position, and derive an analytic perturbative expression that measures the deviation from  the standard fluctuation relations. In Sections 3 and 4, we present the cases of transport in an inhomogeneous medium and in a confined system. Finally, in the last Section, we present our main conclusions.

\section{Diffusion in heterogeneous systems}
To show how Eq.(\ref{cnst-force}) is obtained, let us consider a particle moving in a homogeneous medium subjected to a constant force acting on the $x$-direction, as provided e.g. by gravity~\cite{Astumian_Am_Phys}, an optical trap~\cite{Astumian_JCP} or an entropic force~\cite{david-miguel}. The particle is initially at position $x=0$. The homogeneous nature of the medium leads to a constant diffusion coefficient, $D_0$. Therefore, it is enough to analyze the particle displacement distribution in the direction of the applied force.
In the overdamped regime, the particle dynamics  is governed by the Smoluchowski equation
\begin{equation}
\tb{
\frac{\partial}{\partial t} p(x,t)=-\frac{\partial}{\partial x}\left[D_0\left(\beta f_{0}p(x,t)-\frac{\partial}{\partial x}p(x,t)\right)\right],
}
\label{Eq.gassu-smoluch}
\end{equation}
whose conditional solution, given the initial condition $p(x,0)=\delta (x-x_0)$, reads
\begin{equation}
p(x,t|x_0,0)=\frac{1}{\sqrt{4D_0\pi t}}e^{-\frac{\left(x-\beta D_0 f_{0}t\right)^{2}}{4D_0t}}
\label{Eq.gauss-sol}
\end{equation}
The corresponding ratio between positive, $\Delta x$, and negative, $-\Delta x$, particle displacements reads
\begin{equation}
\tb{
\frac{p(x_0+\Delta x,t|x_0,0)}{p(x_0-\Delta x,t|x_0,0)}=e^{\beta f_{0}\Delta x}
}
\label{verify}
\end{equation}
The quantity $f_{0}\Delta x$ represents the work done on the particle by the force. 

Let us now consider that particles move under the action of an $x$-dependent force, $f(x)=f_0+f_1(x)$, where $f_1=-\partial_xU_0(x)$ is a potential periodic contribution of period $L$ and zero average. A similar form is assumed for the diffusion coefficient: $D(x)=D_0+D_1(x)$ where $D_1(x)$ is periodic of period $L$ and it is vanishing small once averaged over $L$.
The corresponding Smoluchowski equation is given by:
\begin{equation}
\tb{
\frac{\partial}{\partial t}p(x,t)=-\frac{\partial}{\partial x}\left\{D(x)\left[\beta f(x)p(x,t)-\frac{\partial}{\partial x}p(x,t)\right]\right\}.
}
\label{Eq.x-gassu-smoluch}
\end{equation}
In this case, the solution of Eq.(\ref{Eq.x-gassu-smoluch}) is not Gaussian and Eq.~(\ref{verify}) is no longer fulfilled.

To analyze the symmetry of the probability distribution function, we extend Eq.~(\ref{verify}) by introducing 
\begin{equation}
\tb{
\Gamma(\Delta x,t)=\frac{\int p(x_0+\Delta x,t|x_{0},0)p(x_{0},0)dx_{0}}{\int p(x_0-\Delta x,t|x_{0},0)p(x_{0},0)dx_{0}}.
}
\label{Eq.fluct-rel}
\end{equation}
\tb{where $p(x_{0},0)$ is the probability distribution for $t=0$.}
In the case of constant force and diffusion coefficient \tb{and for the initial condition $p(x_{0},0)=\delta(x_0)$}, this expression reduces to $\Gamma=\Gamma_0=e^{\beta f_0\Delta x}$\tb{, i.e. to Eq.~(\ref{verify})}.
An estimate of the  changes in $\Gamma$ due to deviations from Gaussianity can be obtained from
\begin{equation}
 \chi(\Delta x,t)= \frac{\Gamma(\Delta x,t)}{\Gamma_0(\Delta x,t)}
\label{chi}
\end{equation}
Since  analytical solutions of the Smoluchowski equation for $x$-dependent forcing and/or diffusion coefficient are in general difficult to obtain, we will assume that 
\begin{equation}
 p(x,t|x_0,0)\simeq p_0(x,t|x_0,0)+p_1(x,t|x_0,0)
\label{Eq.approx-FJ-sol}
\end{equation}
where $p_0(x,t|x_0,0)$ is given by Eq.(\ref{Eq.gauss-sol}) and $p_1$ is a perturbation. Accordingly, 
\begin{equation}
\tb{
 \int p(x_0+\Delta x,t|x_0,0) p(x_0,0) dx_0 \simeq \rho_0(\Delta x,t)+\rho_1(\Delta x,t)
 }
\label{Eq.approx-FJ-int}
\end{equation}
Substituting  Eq.(\ref{Eq.approx-FJ-int}) in Eq.(\ref{Eq.fluct-rel}), and expanding up to the first order in $\rho_1$ we get
\begin{equation}
 \Gamma(\Delta x,t)\simeq \frac{\rho_0(\Delta x,t)}{\rho_0(-\Delta x,t)}\left[1-\frac{\rho_1(-\Delta x,t)}{\rho_0(-\Delta x,t)} +\frac{\rho_1(\Delta x,t)}{\rho_0(\Delta x,t)} \right]
\label{Gamma-approx}
\end{equation}
therefore $\chi$ reduces to
\begin{equation}
 \chi(\Delta x,t)\simeq1-\frac{\rho_1(-\Delta x,t)}{\rho_0(-\Delta x,t)}+\frac{\rho_1(\Delta x,t)}{\rho_0(\Delta x,t)}.
\label{chi-approx}
\end{equation}
Symmetry enforces that\footnote{We note that $\int_{-\infty}^{\infty} [f(x)-f(-x)]dx=0$}, $\langle \chi\rangle(t)=1/\Lambda\int_{-\Lambda/2}^{\Lambda/2} \chi(x,\Delta x) d\Delta x=1$ from Eq.(\ref{chi-approx}).
It is useful to  consider its averaged second moment
\begin{equation}
 \Omega(t)=\frac{1}{\Lambda}\int_{-\Lambda/2}^{\Lambda/2} \left[\chi(\Delta x,t)-\langle\chi\rangle(t)\right]^2d\Delta x.
\label{omega}
\end{equation}
where $\Lambda$ is the subset over which  $\Omega(t)$ and $\langle \chi\rangle(t)$ are computed\footnote{The choice of $\Lambda$ does not affect significantly the value of $\Omega(t)$ and $\langle \chi\rangle(t)$ and their dependence on $\Lambda$ becomes vanishing small at long time intervals.}. $\Omega$ quantifies the deviations of $\Gamma$ from the homogeneous case for which Eq.~(\ref{verify}) holds. For homogeneous systems, for which $f_1=D_1=0$, one has $\Omega=0$, then \tb{$\Gamma=e^{\beta f_0\Delta x}$, i.e. it recovers the expression in Eq.~(\ref{verify})}.

For both small local forcing, $f_1(x) \ll f_0$, and small modulations of the diffusion coefficient, $D_1(x)\ll D_0$, we can compute $\Omega(t)$ by using the expressions $\rho_0(-\Delta x,t)=\rho_0(\Delta x,t)e^{-\beta f_0 \Delta x}$ and $\rho_1(-\Delta x,t)=\rho_1(\Delta x,t)e^{-\beta \Delta G}$. Expanding  Eq.(\ref{omega})  to first order in $|\beta f_0 \Delta x-\beta \Delta G|\ll1$, one obtains
\begin{eqnarray}
\Omega(t) &\simeq A(t)\langle f_1^2\rangle^2 + B(t)\langle D_1^2\rangle^2+C(t)\langle f_1^2\rangle\langle D_1^2\rangle+\nonumber\\
&+E(t)\langle f_1^2\rangle^\frac{3}{2}\langle D_1^2\rangle^\frac{1}{2}+F(t)\langle f_1^2\rangle^\frac{1}{2}\langle D_1^2\rangle^\frac{3}{2}, \,\,\,\,\,
\label{omega-final}
\end{eqnarray}
where $\langle a(x)\rangle=\frac{1}{\Lambda}\int_{-\Lambda/2}^{\Lambda/2} a(x)dx$ and the time-dependent coefficients $A(t)$, $B(t)$, $C(t)$, $E(t)$ and $F(t)$ are integration constants whose explicit forms are given in the Appendix.

$\Omega$ depends in general on the second moment of the force, $\langle f_1^2\rangle$, and on $\langle D_1^2\rangle$. Hence, to lowest order in both quantities, different physical mechanisms leading to comparable modulations may lead to similar values of $\Omega$.
As shown in the Appendix, the coefficients of the second moment of the forcing, $A(t)$, and of the diffusion coefficient, $B(t)$, are positive, while the  cross terms like $C(t)$, $E(t)$ and $F(t)$ can be positive or negative. Therefore, in the presence of both modulations, the deviations from the Gaussian behavior, that modulate the magnitude of $\Omega$, can either increase or decrease. $\Omega$ also depends implicitly on the average force, $f_0$, through the time-dependent coefficients. 

\section{Results}

In order to study the accuracy of the  perturbative expression Eq.~(\ref{omega-final}), we will consider  two scenarios  where different physical mechanisms lead to a local force and diffusion coefficient. In the first example, we will study the diffusion of particles in an inhomogeneous medium under the influence of a constant force. This case is frequently observed in colloidal suspensions in which particles interact through direct or hydrodynamic interactions and in diffusion in complex systems~\cite{ignacio1994,Hoefling2013,Maes2015,Oshanin2015,Marconi2015}. As a second case, we will analyze the motion of a Brownian particle moving in a confined medium which induces $x$-dependent entropic forces~\cite{jacobs,Zwanzig,Reguera2001,Percus,Kalinay2010,Dagdug,frontiers,Dagdug2013,Kalinay2013,Kalinay2016}. Such a situation is typically observed in molecules moving through ion-channels or membrane pores~\cite{Chinappi2006,Umberto2013,Malgaretti2014,Malgaretti_macromolecules,Malgaretti2015,Bianco2016} and for molecular motors in porous media~\cite{PRE,JCP,MalgarettiEPJST} just to mention a few among others. 

\subsection{Diffusion in an inhomogeneous unbounded medium}
We consider the motion of a Brownian particle moving under the action of a constant force in a medium characterized by a spatially varying diffusion coefficient
\begin{equation}
 D(x)=D_0+D_1\sin\left(2\pi\frac{x}{L}\right)
\label{x-diff}
\end{equation}
The corresponding Smoluchowski equation reads
\begin{equation}
\tb{
\frac{\partial}{\partial t}p(x,t)=-\frac{\partial}{\partial x}\left[D(x)\beta p(x,t)f_0 -D(x)\frac{\partial}{\partial x} p(x,t)\right].
}
\label{Eq.smoluch-diff}
\end{equation}
We have solved Eq.(\ref{Eq.smoluch-diff}) numerically, by means of a Lax-Wendroff method,  with initial condition $p(x,0)=\delta(x)$ \tb{and over a channel made by $10$ identical units each of which is periodic with period $L$
where we have assumed periodic boundary conditions at the channel ends, located at $x=\pm 5L$.}  To avoid the interference of periodic images  we have  followed the evolution of the  particle displacement probability up to a maximum time $T_{max}$ defined as the time at which the ratio $\theta=p(\pm 5L,T_{max})/p(0,T_{max})$ between the probability of particles at the system  edges and the corresponding probability in the middle of the channel overcomes a threshold value, i.e. $\theta \le 10^{-10}$. For $t<T_{max}$ the contribution to $p(x)$ from particles at $x\pm 5L$ is negligible.

\begin{figure*}
\includegraphics[scale=0.43]{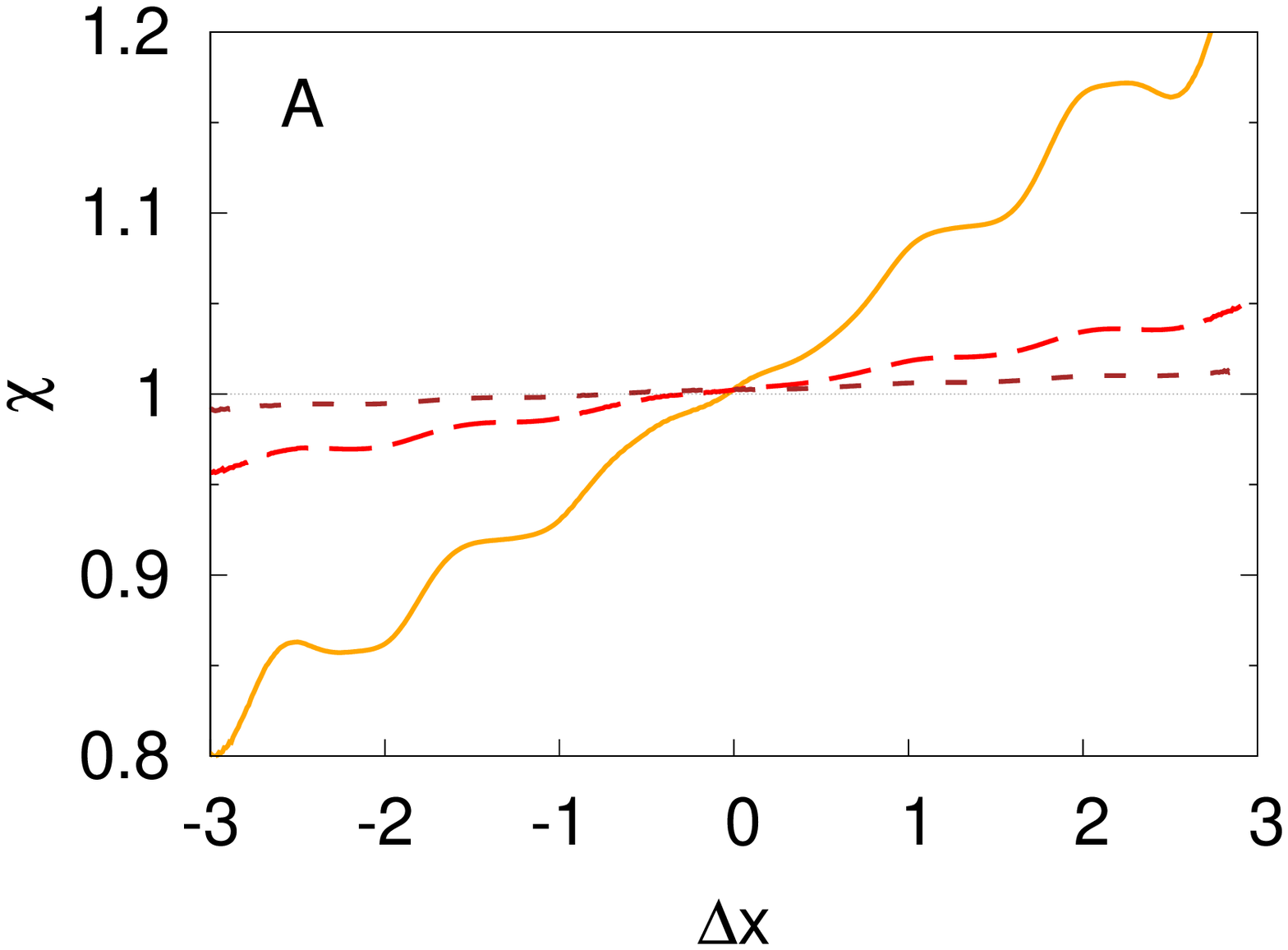}\includegraphics[scale=0.43]{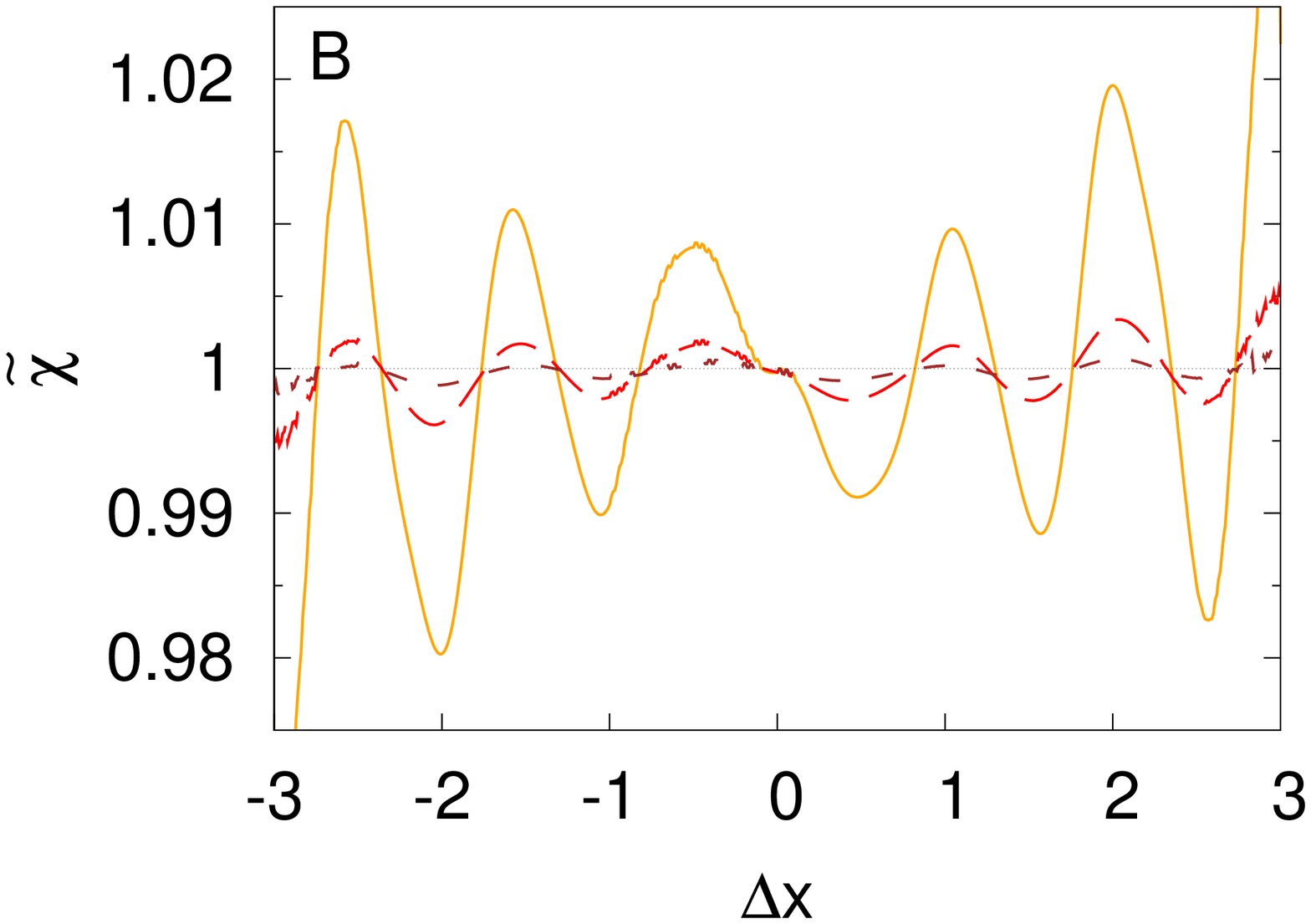}
\includegraphics[scale=0.43]{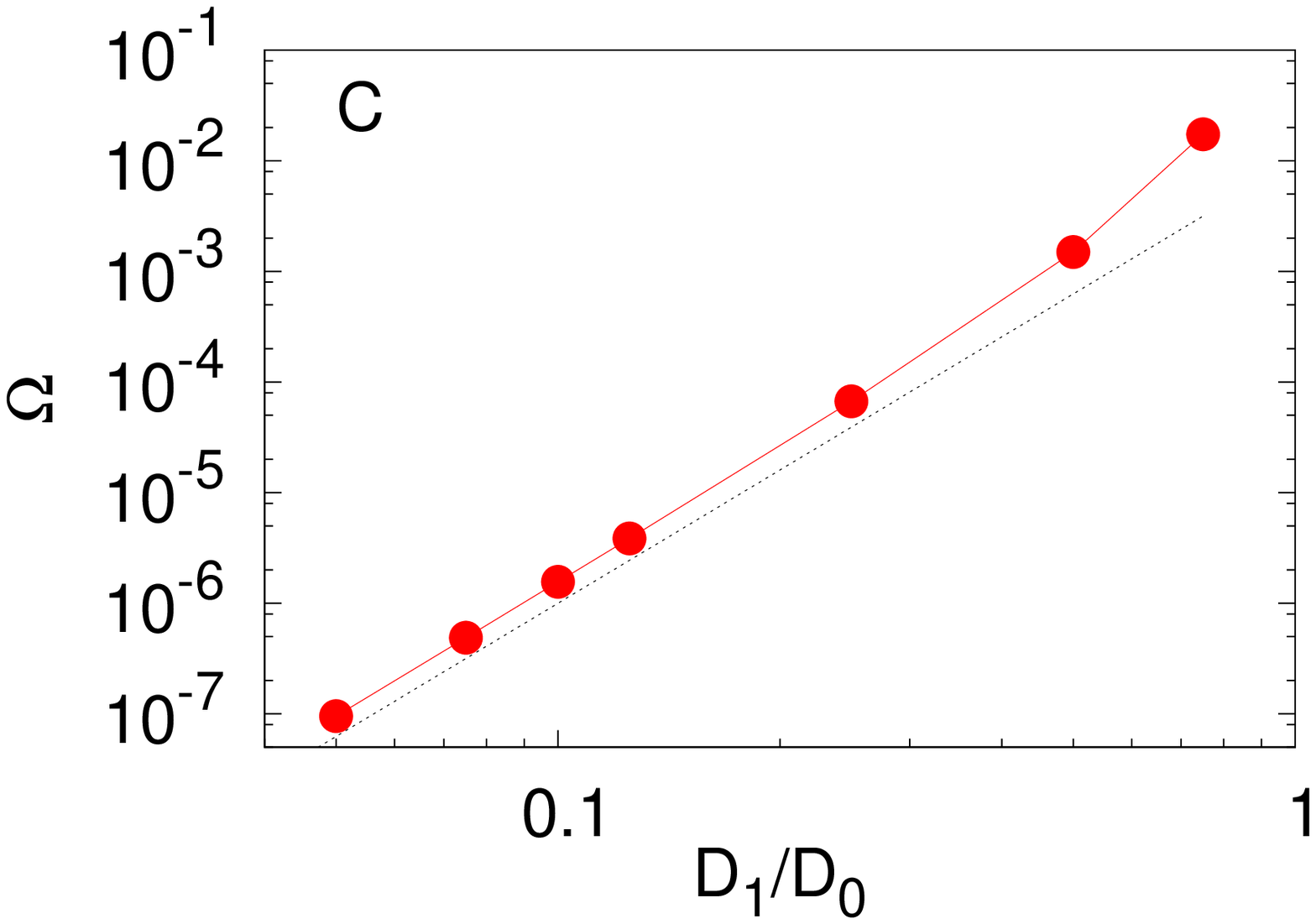}\includegraphics[scale=0.43]{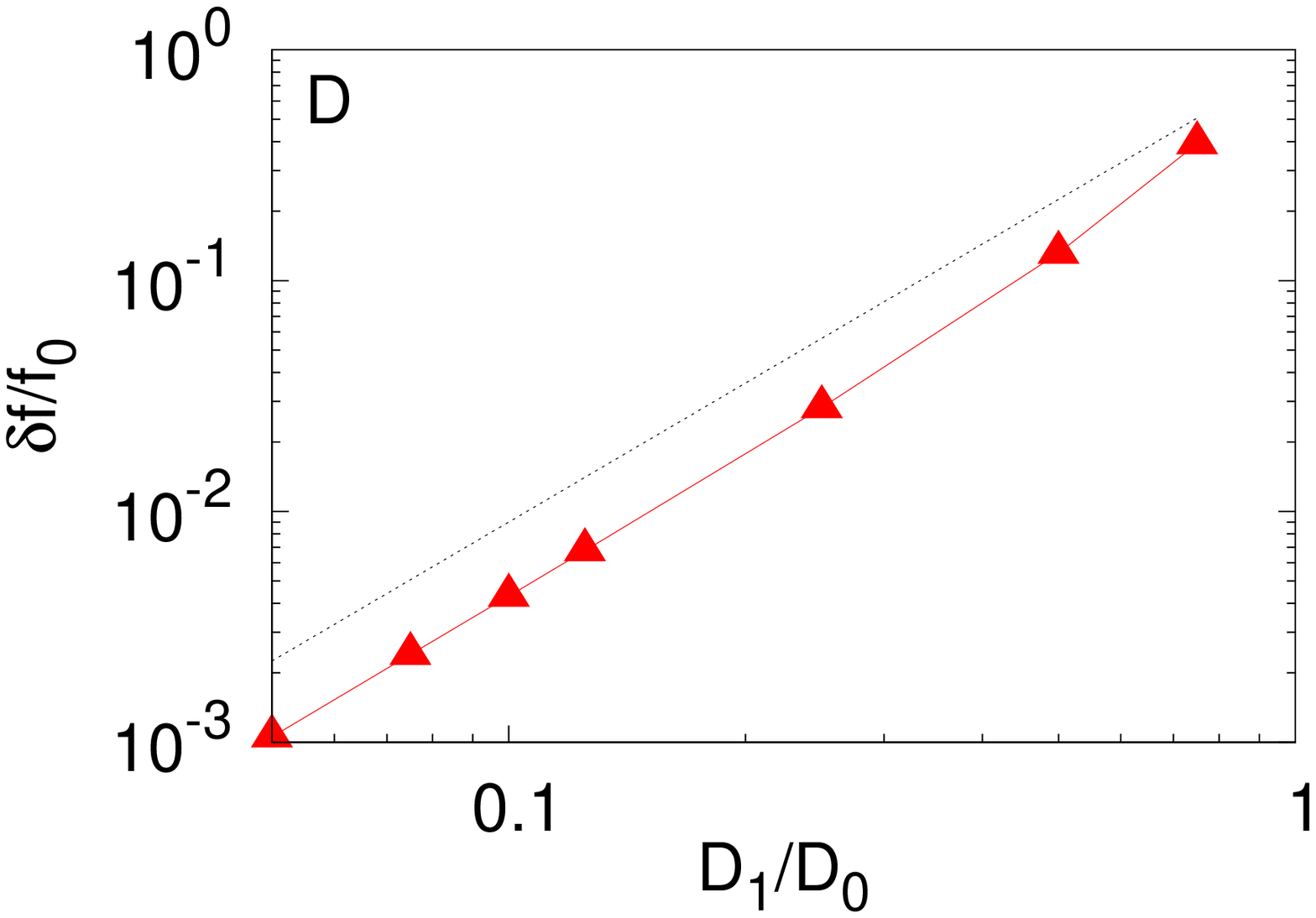}
\includegraphics[scale=0.43]{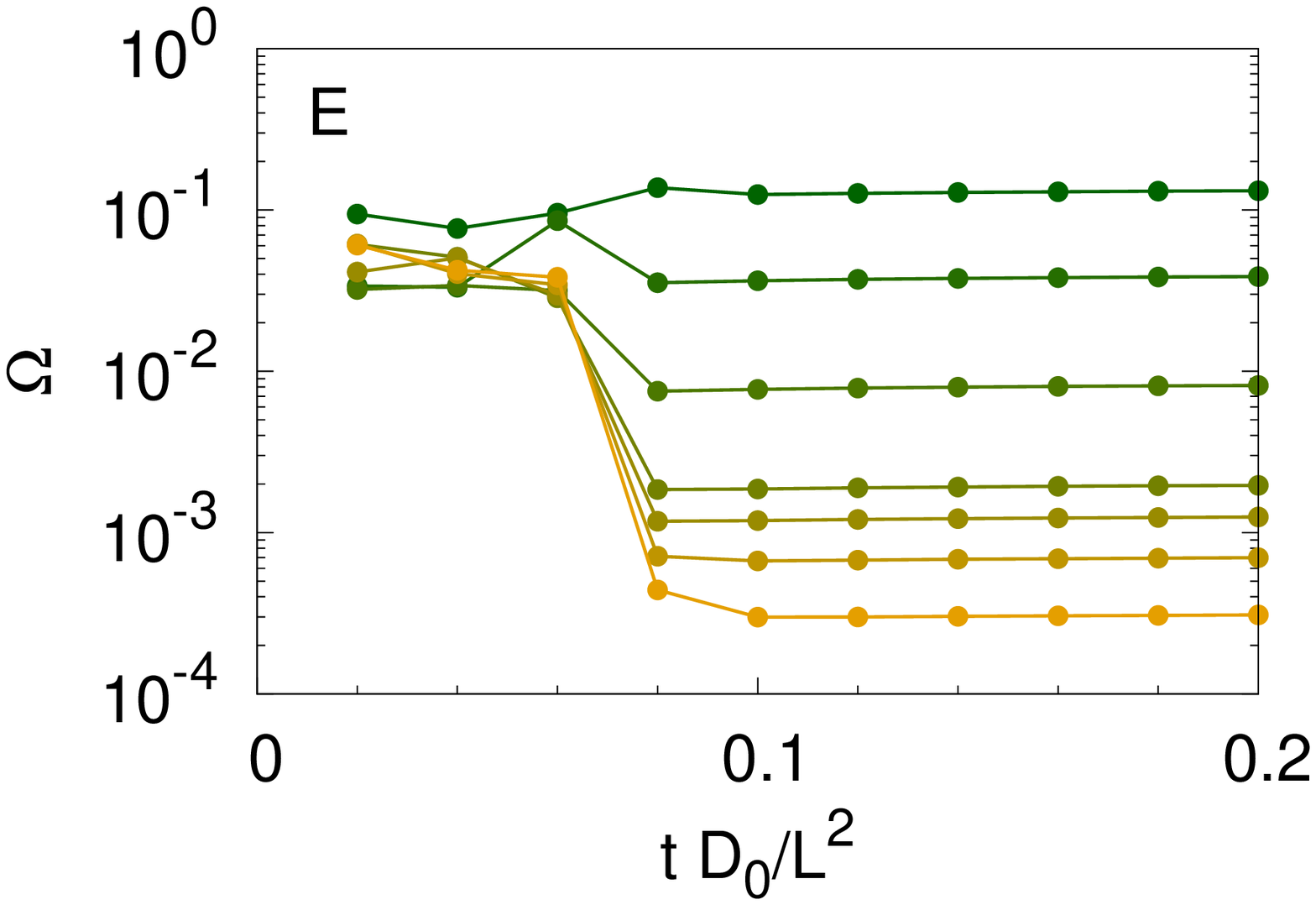}\includegraphics[scale=0.43]{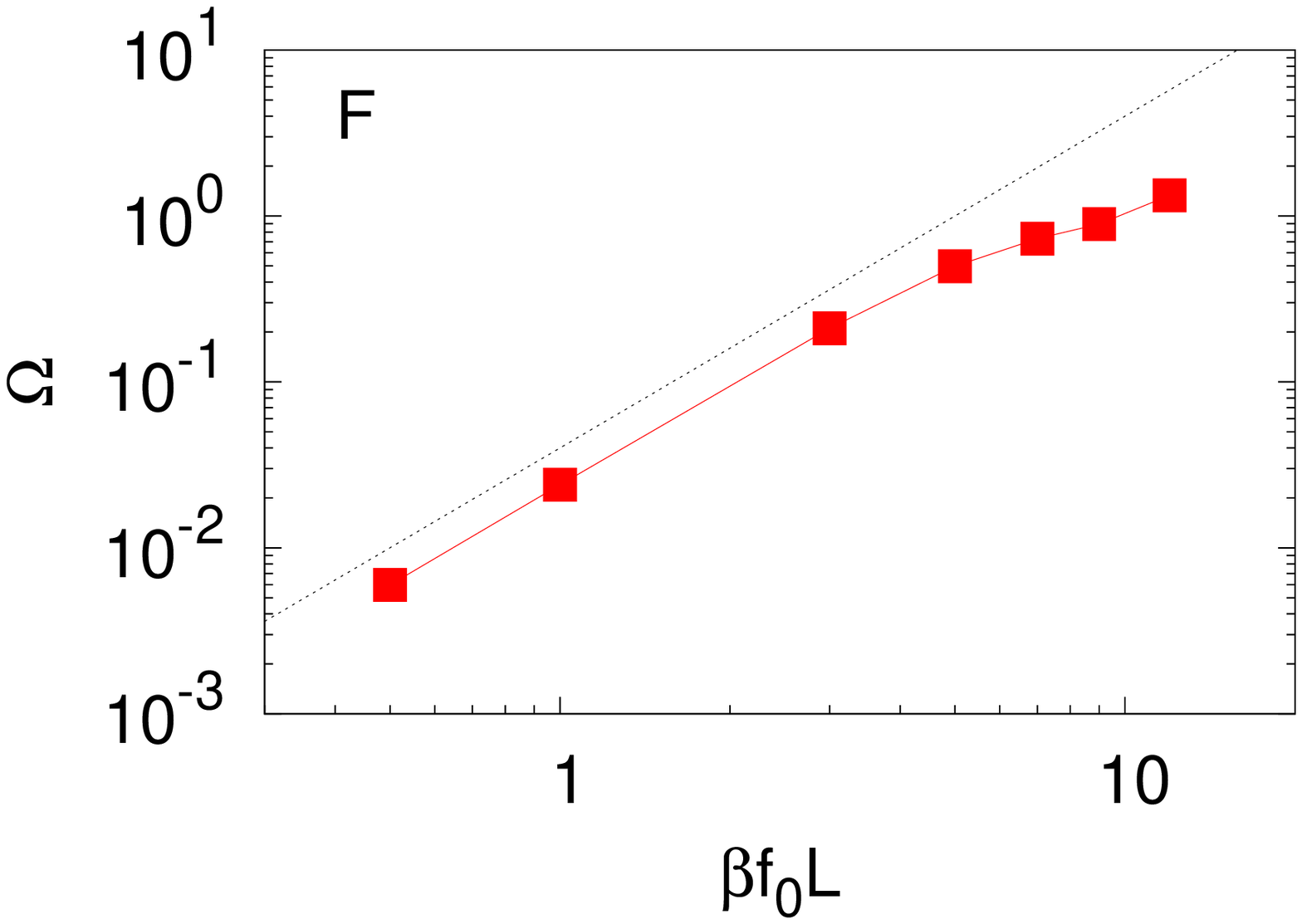}
\caption{Fluctuation relation for a particle in a constant-amplitude channel and inhomogeneous medium, $D_1\ne0$ under a constant force $\beta f_0 L=0.5$. A: $\chi$ as function of the displacement, $\Delta x$, at $t D_0/L^2=0.2$ for different values of the diffusion coefficient modulation: $\frac{D_1}{D_0}=0.125,0.25,0.5$ for dotted (brown), dashed (red) and solid (orange) lines, respectively. B: $\tilde \chi$ as a function of $\Delta x$ for the same parameters as in panel A, where $\delta f$ has been calculated using Eq.(\ref{delta-f}). C: $\Omega$ as a function of $D_1/D_0$ for $\beta f_0 L =0.5$ with $\Lambda=4$ and $t D_0/L^2=0.2$, dotted line stands for $\Omega\propto(D_1/D_0)^4$. D: $\delta f/f_0$ as function of $D_1/D_0$ for the same parameters as in panel C, dotted line stands for $\delta f/f_0\propto (D_1/D_0)^2$. E: $\Omega$ as a function of time normalized by $\tau=L^2/D_0$ with $\Lambda=4$,  for $D_1/D_0=0.05,0.075,0.1,0.25,0.5,0.75$; the darker the line, the larger the ratio $D_1/D_0$. F: $\Omega$ as a function of the driving force $\beta f_0 L$ for $D_1/D_0=0.5$.}
\label{gamma_diff}
\end{figure*}

Fig.(\ref{gamma_diff}.A) shows the dependence of $\chi$ on $ \Delta x$ for different values of $D_1$. For $D_1=0$, $\Gamma$ reduces to Eq.(\ref{verify}) and we recover the expected  relation $\chi=1$. Increasing $D_1$ leads to a non-Gaussian density distribution~\cite{Risken} and  $\chi\ne1$, as shown in Fig.(\ref{gamma_diff}.A). The overall departure from  Gaussianity is  captured better by $\Omega$. As shown in Fig.(\ref{gamma_diff}.C) when increasing the diffusion coefficient modulation, $\Omega$ increases and behaves as $\Omega\propto (D_1/D_0)^4$ in good agreement with Eq.(\ref{omega-final}).
Fig.(\ref{gamma_diff}.A) shows a breakdown of the left-right symmetry superimposed  to a smoother modulation of $\chi$. Indeed, we can regard the system as being driven by an effective force  $f_{eff}=f_0+\delta f$ where $\delta f$ is a  bias. Hence, $\delta f/f_0$ can be regarded as a dimensionless parameter that quantifies particle  rectification arising from 
the interplay between the net force, $f_0$, and the $x$-dependent diffusion coefficient~\footnote{For $\delta f/f_0=0$ no rectification occurs while for $\delta f\neq 0$ the sign of $\delta f$ identifies the rectification direction.}. 

The contribution of $\delta f$ to $\chi$ is given by
\begin{equation}
 \tilde \chi (\Delta x,t)=\chi(\Delta x,t)e^{-\beta \delta f \Delta x},
\end{equation}
that accounts for the deviations from Gaussianity for a system under an effective force. The second moment of $\tilde{\chi}$
\begin{equation}
\tilde\Omega(t) = \frac{1}{\Lambda}\int_{-\Lambda/2}^{\Lambda/2} \left[\tilde \chi(\Delta x,t)-\langle\tilde \chi\right\rangle(t)]^2 d\Delta x
\label{renorm-omega}
\end{equation}
quantifies both the overall departure from  Gaussianity and also provides a route to  obtain $\delta f$. The value of $\delta f$ that better captures the  breakdown of left-right symmetry in Fig.(\ref{gamma_diff}.A) can be obtained by minimizing $\tilde \Omega$. Hence minimizing Eq.~(\ref{renorm-omega}) leads to the following expression for $\delta f$:
\begin{equation}
 \beta \delta f(t)= \frac{\int \chi(\Delta x,t)\Delta x\left[\chi(\Delta x,t)-\langle\chi\rangle(t)\right] d\Delta x}{\int \chi(\Delta x,t)\Delta x^2\left[2\chi(\Delta x,t)-\langle\chi\rangle(t)\right] d\Delta x-\langle \Delta x \chi \rangle^2}.
\label{delta-f}
\end{equation}
Interestingly, Eq.(\ref{delta-f}) predicts $\delta f=0$ for $\chi-\langle \chi \rangle=0$, implying $\Omega=0$. Therefore, in the present regime, no rectification occurs either at equilibrium or for systems leading to a Gaussian distribution of particle displacements.
For vanishing values of $\langle D_1\rangle$ and $\langle f_1\rangle$ we can approximate $\chi(\Delta x,t)\simeq\langle \chi\rangle(t)+\xi\Delta x$, implying $\Omega\propto \xi^2$. In the limit of $\xi\rightarrow 0$, Eq.(\ref{delta-f}) reduces to 
\begin{equation}
 \beta \delta f(t)\simeq \frac{\xi}{\langle \chi\rangle(t)}\propto \sqrt{\Omega}.
\label{delta-f-approx}
\end{equation}
Fig.(\ref{gamma_diff}.B)  displays the dependence of $\tilde\chi$ on $\delta f$, showing the absence of any net tilt.  Therefore, the linear approximation for $\delta f$ given by Eq.(\ref{delta-f}) properly captures the departure of $\chi$, and consequently of $\Omega$, with respect to their values obtained for homogeneous diffusion, $D_1=0$.
Fig.(\ref{gamma_diff}.D) shows the dependence of $\delta f/f_0$ on the modulation in the diffusion. While for larger values of $D_1$ a steeper dependence is observed, for smaller modulations of the diffusion $\delta f/f_0$ reaches an asymptotic behavior $\delta f/f_0\propto D_1^2$.   Comparing the dependence of $\delta f$ and $\Omega$ on $D_1$, we notice that $\delta f\propto \sqrt{\Omega}$, as predicted by Eq.(\ref{delta-f-approx}). 
The regime of validity of Eq.(\ref{omega-final}) is captured in Fig.(\ref{gamma_diff}.C), where the good agreement with the numerical solution of Eq.(\ref{Eq.smoluch-diff}) highlights the wide range of reliability of Eq.(\ref{omega-final}). The temporal evolution of $\Omega$  is shown in Fig.(\ref{gamma_diff}.E). At short times, $\Omega$ displays a remarkable dependence on time and reaches a plateau at longer times, for $t \geq t_0\simeq 0.1 L^2/D_0$. Since $L < \Lambda$,  $\Omega$ relaxes to its steady value faster than particle diffusion over the relevant length scale, $\Lambda$. Finally, Fig.(\ref{gamma_diff}.F) shows the dependence of $\Omega$ on the external constant force, $f_0$, obtaining a quadratic dependence $\Omega\propto f_0^2$ and consequently a linear dependence of $\delta f$ on $f_0$ (data not shown).

\subsection{Diffusion in a periodic channel}

We consider the diffusion of a particle in a channel of periodic half-section
\begin{equation}
h(x)=h_{0}+h_{1}\cos\left(2\pi\frac{x}{L}\right),\label{Eq.channel-shape}
\end{equation}
where $L$ is the period and $L_z$ the width along the $z$-direction assumed to be constant.
In the overdamped regime, the evolution of the probability density function, $P(x,y,z,t)$, of the particle under the action of a constant force, $f_0$, is governed by the $3D$ Smoluchowski equation: 
\begin{equation}
\tb{
\frac{\partial}{\partial t}P(x,y,z,t)=D\beta\nabla\cdot\left[P(x,y,z,t)\nabla U(x,y,z)+D\nabla P(x,y,z,t)\right]
}
\label{Eq.smoluch-channel}
\end{equation}
where  the potential $U(x,y,z)$ is given by
\begin{eqnarray*}
U(x,y,z) & = & U(x+L,y,z)\nonumber\\
U(x,y,x)&=&\begin{cases} 
f_0 x, \,|y|\le h(x)\,\&\,|z|\le L_{z}/2\\
\infty,\,\,\,\,\,\,\,\,|y|>h(x)\, or\,|z|>L_{z}/2
\end{cases}
\end{eqnarray*}
and involves both the external driving, $f_{0}$, and the presence of boundaries. For smoothly varying channel amplitudes, $\partial_{x}h(x)\ll1$,  the diffusing particles equilibrates much faster in the transverse direction than in the main transport direction. One can then assume
\begin{eqnarray}
P(x,y,z,t) & = & p(x,t)\frac{e^{-\beta U(x,y,z)}}{e^{-\beta A(x)}}\\
e^{-\beta A(x)} & = & \int_{L_{z}/2}^{L_{z}/2}\int_{-h(x)}^{h(x)}e^{-\beta U(x,y,z)}dydz,
\end{eqnarray}
where $p(x,t)$ is the probability distribution in the coarse-grained description and $A(x)$ is the corresponding free energy
\begin{equation}
 A(x)=f_{0}x-\frac{1}{\beta}\ln\left[h\left(x\right)\right].
\label{x-free-energy}
\end{equation}
This quantity consists of an enthalpic contribution, $f_{0}x$, and an entropic contribution, $-\frac{1}{\beta}\ln\left[h\left(x\right)\right]$. This approximation shows that diffusion in $3D$ can be analyzed through $1D$ diffusion in the presence of entropic barriers~\cite{jacobs,Zwanzig,Reguera2001,Percus,martens2011entropic}.
Accordingly, we can define the dimensionless energy barrier that the particles experience along the channel,
\begin{equation}
 \Delta S=\ln\left[\frac{h_{max}}{h_{min}}\right]
\label{deltaS}
\end{equation}
where $h_{min}$ and $h_{max}$ are the minimum and maximum channel aperture, respectively.
Integrating Eq.(\ref{Eq.smoluch-channel}) along the channel transverse section, we obtain the Fick-Jacobs equation 
\begin{equation}
\tb{
\frac{\partial}{\partial t}p(x,t)=\frac{\partial}{\partial x}\left\{ D(x)\left[\beta p(x,t)\frac{\partial}{\partial x}A(x)+\frac{\partial}{\partial x}p(x,t)\right]\right\}. 
}
\label{Eq.fick-jacobs}
\end{equation}
where
\begin{equation}
 D(x)=\frac{D_0}{\left[1+\left(\partial_xh(x)\right)^2\right]^\alpha}
\label{diff-approx}
\end{equation}
is an effective diffusion coefficient, with alpha $\alpha=1/3(1/2)$  in three(two) spatial dimensions~\cite{Reguera2001}. 
Comparison of Eq.(\ref{Eq.fick-jacobs}) with Eq.(\ref{Eq.x-gassu-smoluch}) shows that the geometrical confinement enters  through the potential $A(x)$. Its spatial derivative gives rise to an effective force; therefore, we can understand the  impact of the channel corrugation as providing a spatially-varying force acting on the Brownian particle. 
\begin{figure*}
\includegraphics[scale=0.43]{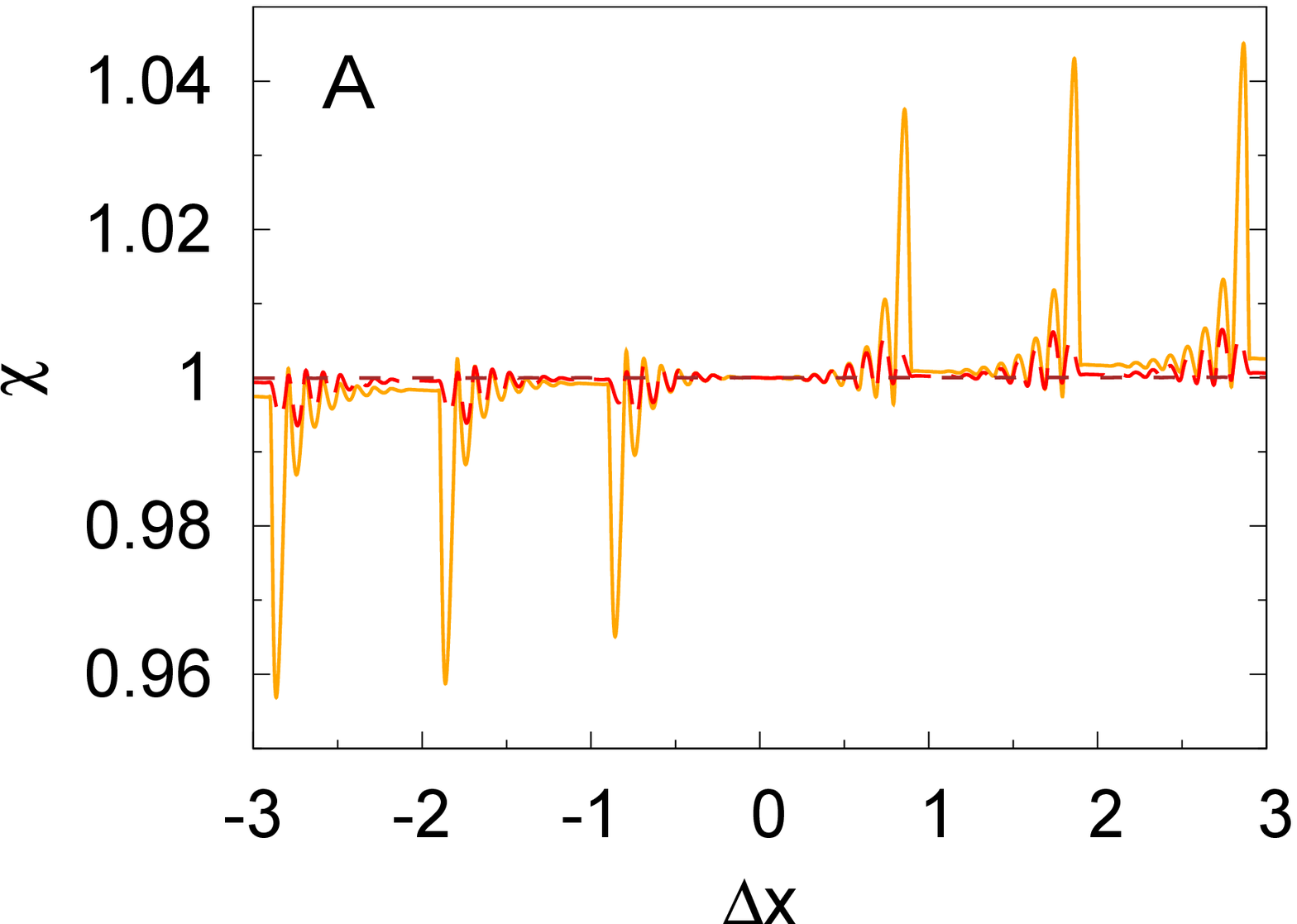}\includegraphics[scale=0.43]{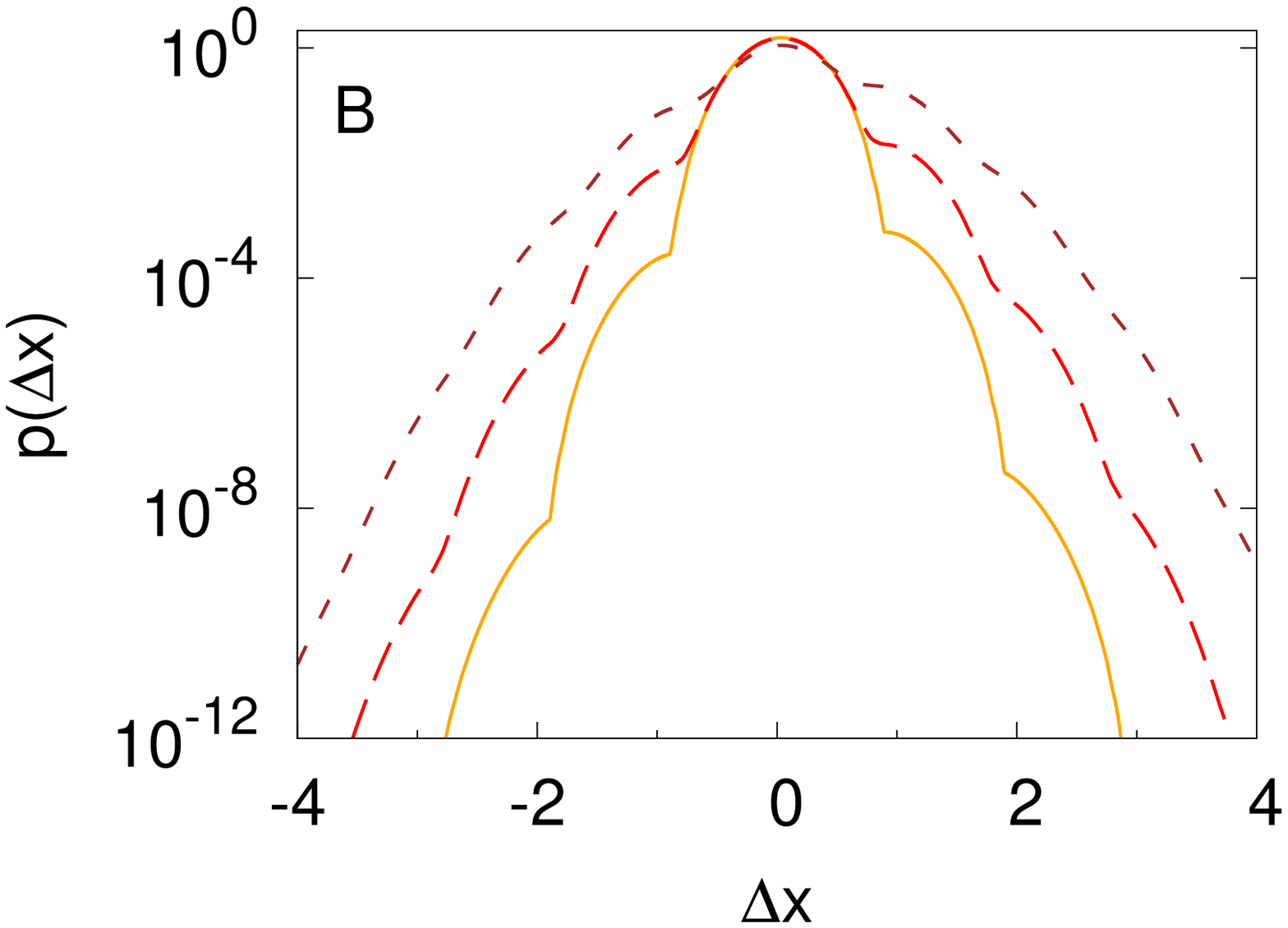}
\includegraphics[scale=0.43]{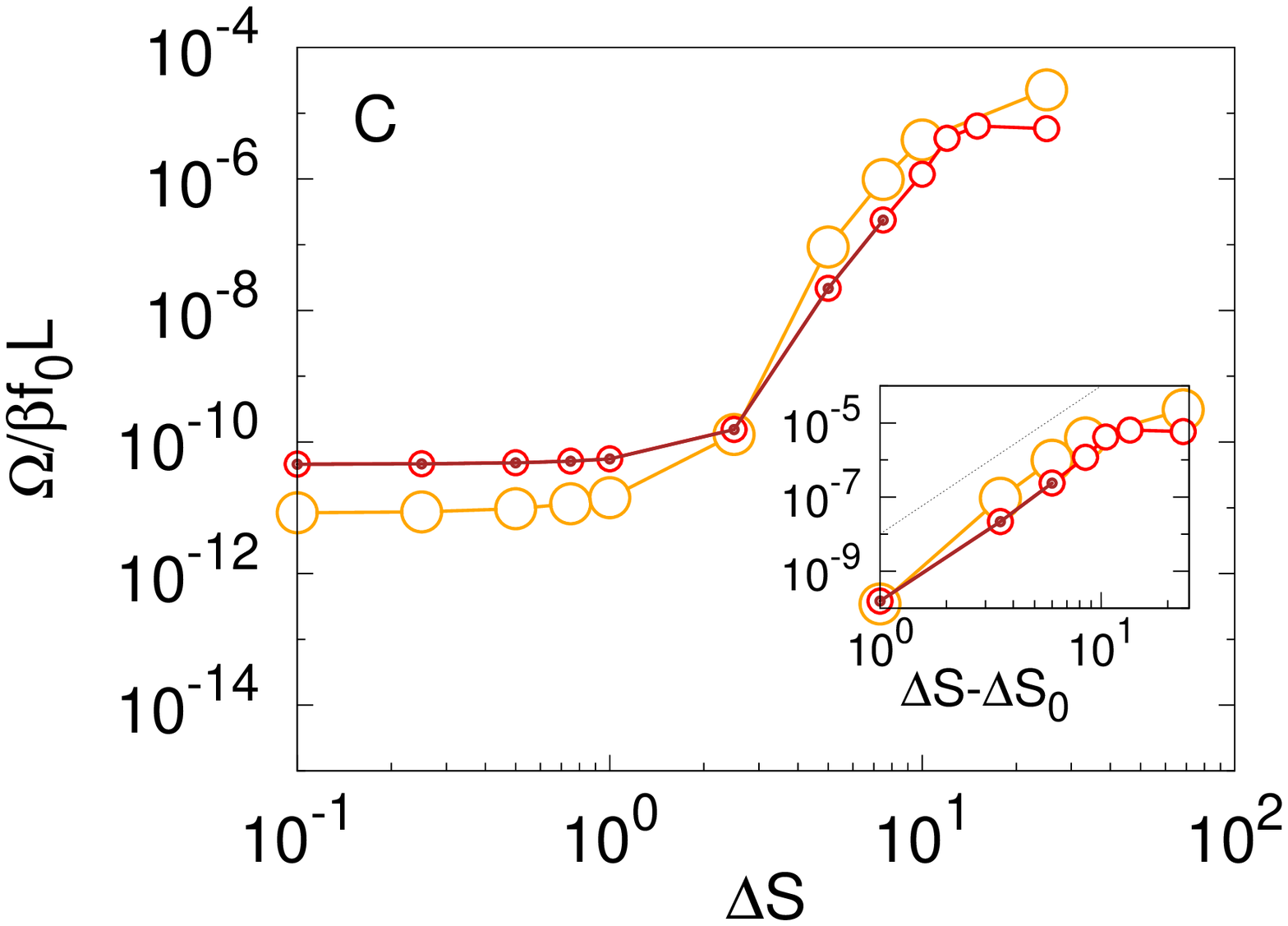}\includegraphics[scale=0.43]{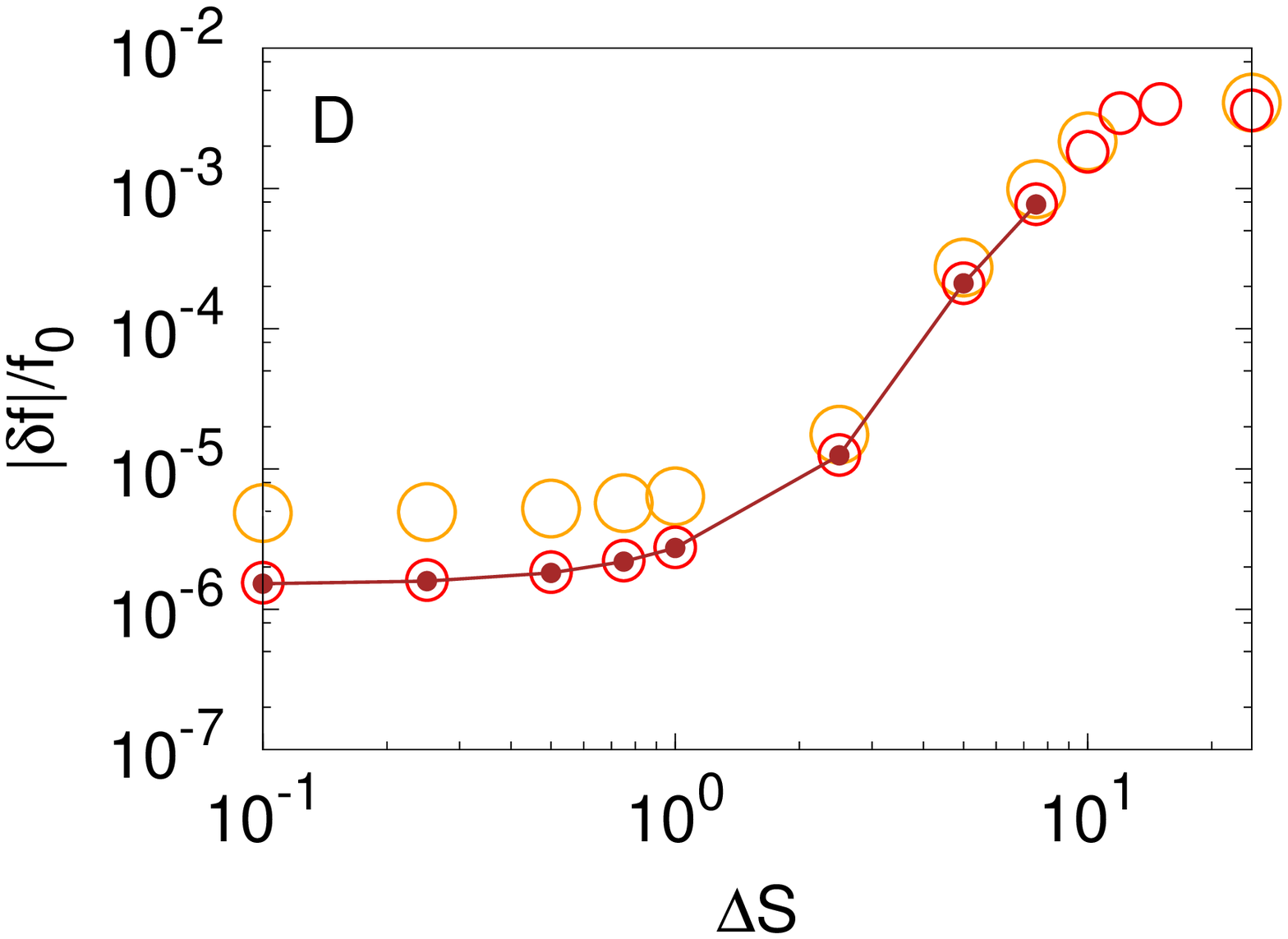}
\includegraphics[scale=0.43]{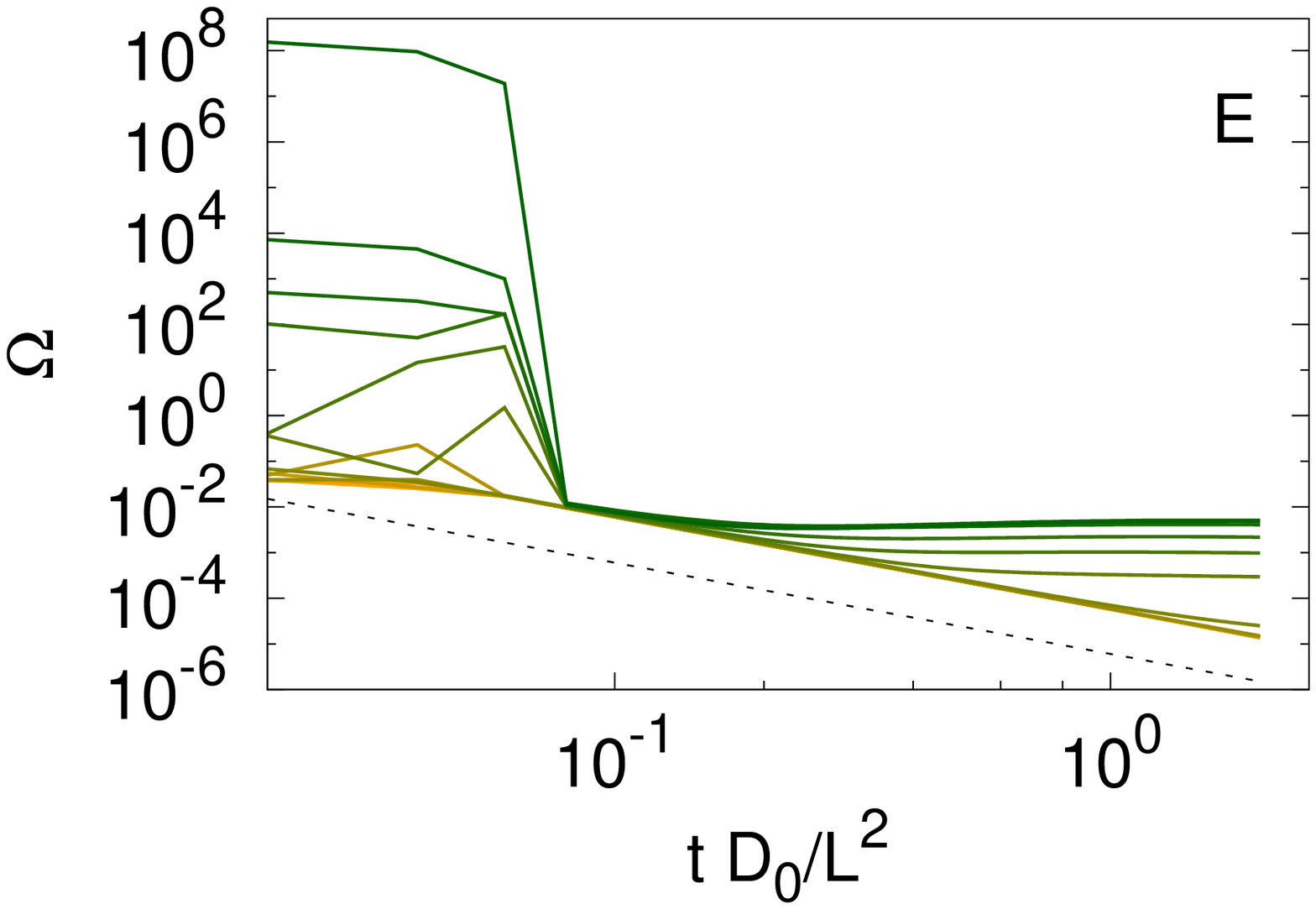}\includegraphics[scale=0.43]{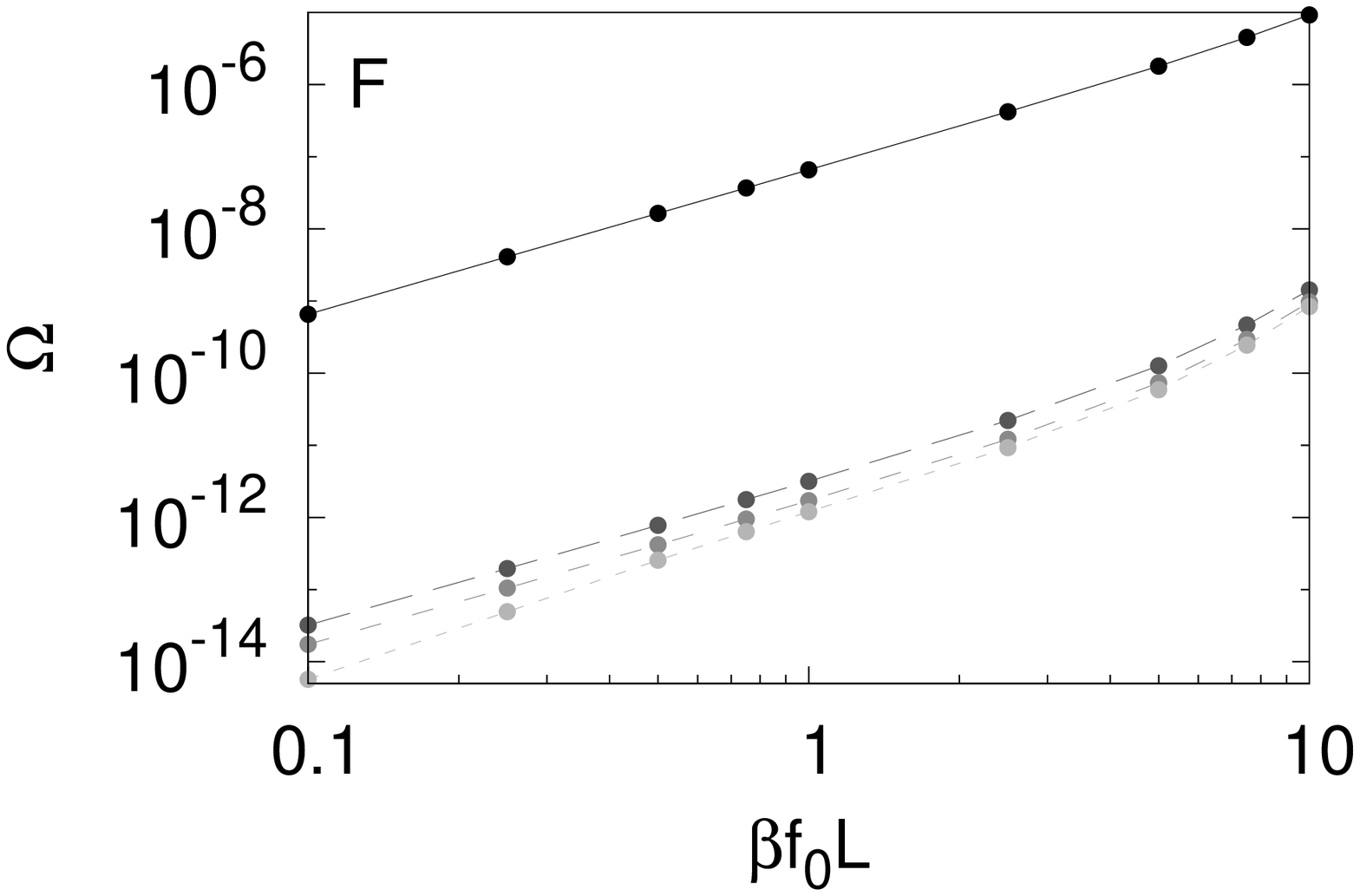}
\caption{Fluctuation relation for a Brownian particle in a varying-amplitude channel, $h_1\neq 0$, that induces a modulation in the effective diffusion coefficient, $D_1\ne0$ \tb{for} a constant force $\beta f_0 L$. A: $\chi$ as function of the displacement, $\Delta x$, \tb{for} $t D_0/L^2=2$, with $\beta f_0 L=1$ and different values of the entropic barrier $\Delta S=0.1,1,10$ for dotted (brown), dashed (red) and solid (orange) line respectively. B: $p(\Delta x)$ as function of $\Delta x$ for \tb{the same values of the parameters} as in panel A. C: $\Omega$, at $t D_0/L^2=2$, as a function of \tb{$\Delta S$} for $\beta f_0 L =0.1,1,10$ standing bigger points for larger $\Delta S$, with $\Lambda=4$. Inset: $\Omega$ as a function of the reduce \tb{entropic} barrier $\Delta S-\Delta S_0$ with $\Delta S_0=1.5$, for the same value of the parameters in the main figure; dotted line stands for $\Omega\propto(\Delta S-\Delta S_0)^4$. D: $|\delta f|/f_0$ as function of $\Delta S$ for the same value of the parameters in panel C. E: $\Omega$ as a function of 
time normalized by $\tau=L^2/D_0$ for $\Delta S=0.1,0.25,0.5,0.75,1,2.5,5,7.5,10$, standing darker lines for larger values of $\Delta S$. F: $\Omega$ as a function of the forcing $\beta f_0 L$ for $\Delta S=0.1,0.5,1,5$ standing darker lines for higher values of $\Delta S$, being $\Lambda=4$. }
\label{gamma_fick}
\end{figure*}

We have solved numerically \tb{Eq.(\ref{Eq.fick-jacobs})} with the same numerical scheme used in the previous section. Fig.(\ref{gamma_fick}.A) shows the behavior of $\chi$. Analogously to the  results reported in the previous section, $\chi$ is strongly affected by the local drift and diffusion coefficient modulation. Larger values of $\Delta S$, i.e. larger modulations, lead to a more involved dependence of $\chi$ on $\Delta x$, and  consequently to a larger departure from the Gaussian solution obtained for $\Delta S=0$. Moreover, comparing Fig.(\ref{gamma_fick}.A) and Fig.(\ref{gamma_diff}.A), we notice that the qualitative and quantitative behaviors of $\chi$ differ for bounded and unbounded diffusion.
While the latter case is characterized by a smoothly modulated overall extra-tilt for $\chi$, in the former larger modulations are overimposed to a smoothly-varying tilt, even for entropy barriers as large as $\Delta S=10$.

The dependence of $\Omega$ on $\Delta S$ is also modified with respect to the behavior observed for a constant channel section, as can be appreciated comparing  Figs.(\ref{gamma_diff}.C) with Fig.(\ref{gamma_fick}.C). In the system analyzed, $\Omega$ shows a weaker dependence on $\Delta S$. Disentangling the underlying mechanisms responsible for this lack of sensitivity is not straightforward, because modulations of $D_1$ and $f_1$ due to   variations of the channel section compete  with each other, as becomes clear in Eq.(\ref{omega-final}). Nonetheless, for $\Delta S>1$, using Eq.(\ref{diff-approx}), we have
\begin{equation}
 \langle D_1^2\rangle\propto \frac{e^{\Delta S}-1}{e^{\Delta S}+1}
\end{equation}
implying that $\langle D_1^2\rangle$ vary very smoothly for larger $\Delta S$. Thus, the  dependence of $\Omega$ on $\Delta S$ enters essentially through the entropic force. One can assume that $\langle D_1^2\rangle$ is practically constant to obtain:
\begin{equation}
 \Omega\simeq (\Delta S-\Delta S_0)^4
\label{delta_S0}
\end{equation}
where $\Delta S_0$ accounts for the contribution coming from the modulation in the diffusion coefficient.
The inset of Fig.(\ref{gamma_fick}.C) shows the good agreement of the theoretical prediction with the numerical results, up to $\Delta S \simeq 10$.
The deviation from the behavior $\Omega\propto (\Delta S-\Delta S_0)^4$  observed for smaller values of $\Delta S$ is due to the time-dependence of $\Omega$. As shown in Fig.(\ref{gamma_fick}.E), $\Omega$ reaches a quasi-steady state after a transient that depends on  $\Delta S$. For increasing entropy barriers, $\Delta S$, the effective forces acting on a Brownian particle increase leading to a reduction of the relaxation time, as it happens for particles in a  potential well~\cite{Risken}.  Smaller values of $\Delta S$ require longer relaxation times that cannot be considered in our numerical solution.

Even though the dependence of $\chi$ on $\Delta x$ is quite involved and does not show a clear breaking of the left-right symmetry, we have used Eq.(\ref{delta-f}) to compute the rectification parameter $\delta f$. It results that $\delta f\propto (\Delta S-\Delta S_0)^4$ as shown in Fig.(\ref{gamma_fick}.D). Since $\Omega\propto (\Delta S-\Delta S_0)^4$, we predict $\delta f \propto \Omega$, that differs from the behavior observed in the previous case in which variations of $\chi$ led to $\delta f\propto\sqrt{\Omega}$. 

We can then conclude that different  local transport mechanisms  lead to different relationships between the rectification parameter and the deviations from Gaussianity inherent to Omega. 
Fig.(\ref{gamma_fick}.F) displays the  dependence of $\Omega$ on the external force and shows that for decreasing forces the deviation in $\Omega$ become vanishing small recovering  the equilibrium value $\Omega=0$, for $\beta f_0 L=0$.  Moreover, $\Omega \sim f_0^2$  (Fig.(\ref{gamma_fick}.F)), as also observed for a constant section channel (Fig.(\ref{gamma_diff}.F)).

\section{Discussion}

We have shown that the diffusion of particles is strongly affected by heterogeneities resulting from irregularities of the boundaries or from the intrinsic nature of the host medium. 
The presence of local forces or of a local diffusion coefficient breaks down the Gaussian form of the probability distribution for the particles and leads to an effective rectification.

For small modulations of the spatial heterogeneities it is possible to analyze the consequences of a non-Gaussian probability distribution. We have found that the ratio between the probabilities of  forward and backward moves depends on the heterogeneities of the medium and also on time. We have derived an expression for their ratio, $\Gamma$ (Eq.(\ref{Eq.fluct-rel})), that is valid for  small modulations both in the forcing and/or in the diffusion coefficient. In order to quantify the average deviation from Gaussian behavior, higher moments of $\Gamma$ are insightful. The functional shape of the second moment $\Omega$ of $\Gamma$ shows that the corrections to $\Gamma$, induced by local transport are proportional to the dispersion of the modulation. When both force and diffusion coefficient are modulated, Eq.(\ref{omega-final}) predicts that different regimes can be achieved depending on the constructive or destructive interaction between the two mechanisms. 
 
To test our predictions, we have checked  Eq.(\ref{omega-final})  for two different scenarios, namely a particle moving in an inhomogeneous medium with a position-dependent diffusion coefficient  and a particle in a channel of varying cross section in the presence of entropic forces. In the first case, the force exerted is  constant whereas the diffusion coefficient depends on position. In the latter case, both the geometrically-induced effective  force and the local particle diffusion coefficient depend on particle's position along the channel.
In both situations we observe a remarkable agreement between the numerical results and our prediction for $\Omega$ in the case of mild variations in the forcing and/or medium heterogeneities. 

The coupling between local forcing and diffusion can also lead to particle rectification; our analysis predicts when rectification emerges and identifies an effective parameter, $\delta f$, that quantifies the effective rectification. In particular, our analysis reveals how the dependence of rectification on the departure from Gaussianity  is affected by the physical mechanism responsible for  local transport. 
These results suggest a possible way to characterize the intrinsic properties of the host medium and of the confinement based on the use of the new fluctuation relation and the tracking of the particles. 

\vspace{6pt}

\section*{Acknowledgments}
We acknowledge MINECO and DURSI  for financial support
under projects  FIS\ 2015-67837-P and 2014SGR-922, respectively. J.M. Rubi and I. Pagonabarra acknowledges financial support from {\sl Generalitat de Catalunya } under program {\sl Icrea Academia}. P.M. thanks Marco Ribezzi for useful discussions.

\appendix
\section{}
Here we derive the expression for $\Omega$ given by Eq.(\ref{omega-final}). 
Substituting Eq.(\ref{Gamma-approx}) into Eq.(\ref{omega}) and remembering that $\langle\chi\rangle=1$ when calculate by Eq.(\ref{chi-approx}) we obtain:
\begin{equation}
\Omega=\int \left[ \frac{\rho_1(-\Delta x,t)}{\rho_0(-\Delta x,t)}-\frac{\rho_1(\Delta x,t)}{\rho_0(\Delta x,t)}\right]^2d\Delta x
\end{equation}
Remembering that $\rho_0(-\Delta x,t)=\rho_0(\Delta x,t)e^{-\beta f_0 \Delta x}$ and assuming $\rho_1(-\Delta x,t)=\rho_1(\Delta x,t)e^{-\beta \Delta G}$ we obtain:
\begin{equation}
 \Omega=\int \left[\frac{\rho_1(\Delta x,t)}{\rho_0(\Delta x,t)}\right]^2\left(e^{\beta f_0 \Delta x-\beta \Delta G}-1\right)d\Delta x.
\end{equation}
Expanding for $|\beta f_0 \Delta x-\Delta G|\ll1$ we get:
\begin{equation}
\Omega=\int \left(\beta f_0\Delta x-\beta \Delta G\right)^2\left[ \frac{\rho_1(\Delta x,t)}{\rho_0(\Delta x,t)}\right]^2d\Delta x
\label{app-omega-last}
\end{equation}
where we have used the fact that $\int \left(\beta f_0\Delta x-\beta \Delta G\right)\left[ \frac{\rho_1(\Delta x,t)}{\rho_0(\Delta x,t)}\right]^2 d\Delta x=0$ due to the even character of $\left[ \frac{\rho_1(\Delta x,t)}{\rho_0(\Delta x,t)}\right]^2$ in the limit $|\beta f_0 \Delta x-\Delta G|\ll 1$ and to the odd character of $\beta f_0\Delta x-\beta \Delta G$. 
Finally we expand both $\beta \Delta G$ and $\rho_1$ as a power series of the local modulation of the forcing, $f_1(x)$, and/or diffusion coefficient, $D_1(x)$. Since both $f_1(x)$ and $D_1(x)$ are periodic with zero mean, see Eq.(\ref{x-diff}),(\ref{x-free-energy}) the first non vanishing contribution to $\beta \Delta G$ and $\rho_1$ is provided their second moment:
\begin{eqnarray}
 \langle f_1^2\rangle=\int f_1^2(x)dx&\\
 \langle D_1^2\rangle=\int D_1^2(x)dx&
\end{eqnarray}
When the modulations are vanishing small we have that $\rho_1\rightarrow 0$ and, remembering  $\rho_1(-\Delta x,t)=\rho_1(\Delta x,t)e^{-\beta \Delta G}$, we have $\Delta G \rightarrow f_0 \Delta x$. 
Using the last expressions we can expand $\Delta G$ and $\rho_1$ up to first order:
\begin{eqnarray}
&\beta \Delta G=\beta f_0 \Delta x+\alpha_1(\Delta x) \sqrt{\langle f_1^2\rangle} + \alpha_2(\Delta x) \sqrt{\langle D_1^2\rangle}\,\,\,\,\,\,\,\,\,\,\,\,\,\,\,\,\, \label{app-Delta}\\
&\rho_1(\Delta x,t)=\gamma_1(\Delta x,t) \sqrt{\langle f_1^2\rangle} + \gamma_2(\Delta x,t) \sqrt{\langle D_1^2\rangle}\,\,\,\,\,\,\,\,\,\,\,\,\,\,\,\,\, \label{app-DeltaRho}
\end{eqnarray}
where $\alpha_1(\Delta x)$, $\alpha_2(\Delta x)$, $\gamma_1(\Delta x,t)$ and $\gamma_2(\Delta x,t)$ are to be determined numerically or by a path integral solution of Eq.(\ref{Eq.fick-jacobs}). Substituting Eq..(\ref{app-Delta}),(\ref{app-DeltaRho}) into Eq.(\ref{app-omega-last}) we get:
\begin{multline}
\Omega=\int \left(\alpha_1(\Delta x,t) \sqrt{\langle f_1^2\rangle} + \alpha_2(\Delta x,t) \sqrt{\langle D_1^2\rangle}\right)^2\cdot\\
\left[\gamma_1(\Delta x,t) \sqrt{\langle f_1^2\rangle} + \gamma_2(\Delta x,t) \sqrt{\langle D_1^2\rangle}\right]^2\frac{d\Delta x}{\rho^{2}_0(\Delta x,t)}
\end{multline}
and finally we can define the coefficient that appears in Eq.(\ref{omega-final}) as:
\begin{eqnarray*}
A(t)&=&\int \alpha_1^2(\Delta x)\gamma_1^2(\Delta x,t)\frac{d\Delta x}{\rho^{2}_0(\Delta x,t)} \nonumber\\
B(t)&=&\int \alpha_2^2(\Delta x)\gamma_2^2(\Delta x,t)\frac{d\Delta x}{\rho^{2}_0(\Delta x,t)}\\
C(t)&=&\int \alpha_1^2(\Delta x)\gamma_2^2(\Delta x,t)+\alpha_2^2(\Delta x)\gamma_1^2(\Delta x,t)+4\alpha_1(\Delta x)\alpha_2(\Delta x)\gamma_1(\Delta x,t)\gamma_2(\Delta x,t) \frac{d\Delta x}{\rho^{2}_0(\Delta x,t)}\nonumber\\
E(t)&=&2\int\left(\alpha_1(\Delta x) \gamma_2(\Delta x,t)+\alpha_2(\Delta x) \gamma_1(\Delta x,t) \right)\alpha_1(\Delta x) \gamma_2(\Delta x,t) \frac{d\Delta x}{\rho^{2}_0(\Delta x,t)}\nonumber\\
F(t)&=&2\int\left(\alpha_1(\Delta x) \gamma_2(\Delta x,t)+\alpha_2(\Delta x) \gamma_1(\Delta x,t) \right)\alpha_2(\Delta x) \gamma_1(\Delta x,t) \frac{d\Delta x}{\rho^{2}_0(\Delta x,t)}\nonumber
\end{eqnarray*}




\bibliography{fluct_jcp_references}

\end{document}